\documentclass{IEEEtran}
\usepackage{cite}
\usepackage{amsmath,amssymb,amsfonts}
\usepackage{graphicx}
\usepackage{textcomp,nicefrac}
\def\BibTeX{{\rm B\kern-.05em{\sc i\kern-.025em b}\kern-.08em
T\kern-.1667em\lower.7ex\hbox{E}\kern-.125emX}}
\begin{document}
\title{Archiver System Management\\ for Belle II Detector Operation}
\author{Y.-K. Kim, S.-J. Cho, S.-H. Park, M. Nakao and T. Konno
\thanks{
Y.-K. Kim, and S.-J.Cho are with the Department of Physics, Yonsei University, Seoul, Republic of Korea. (e-mail: ykjk1401@yonsei.ac.kr, sjcho93@yonsei.ac.kr).}
\thanks{S.-H. Park, was with the Yonsei University, Seoul, Republic of Korea. 
He is now with the KEK, Tsukuba, Japan (e-mail: seokhee.park@yonsei.ac.kr).}
\thanks{M. Nakao is with
Institute of Particle and Nuclear Physics, KEK,
Tsukuba, Japan (e-mail: mikihiko.nakao@kek.jp).}
\thanks{T. Konno is with
Department of Physics, Kitasato University,
Sagamihara, Japan (e-mail: tkonno@kitasato-u.ac.jp).}
}

\maketitle

\begin{abstract}
    The Belle II experiment is a high-energy physics experiment at the SuperKEKB electron-positron collider. 
    Using Belle II data, high precision measurement of rare decays and CP-violation 
    in heavy quarks and leptons can be performed to probe New Physics. 
    In this paper, we present the archiver system used to store the monitoring data of
    the Belle II detector 
    and discuss in particular how we maintain the system that archives the monitoring process variables of the subdetectors. 
    We currently save about 26 thousand variables including the temperature of various subdetectors components, 
    status of water leak sensors, high voltage power supply status, data acquisition status, and luminosity information of the colliding beams. 
    For stable data taking, it is essential to collect and archive these variables.
    We ensure the availability and consistency of all the variables from the subdetector and other systems, 
    as well as the status of the archiver itself are consistent and regularly updated.
    To cope with a possible hardware failure, 
    we prepared a backup archiver that is synchronized with the main archiver.
\end{abstract}

\begin{IEEEkeywords}
    Belle II, Monitoring
\end{IEEEkeywords}

\section{Introduction}
\label{sec:introduction}
\IEEEPARstart{T}{he} Belle II experiment\cite{b1} at the SuperKEKB collider at KEK, Tsukuba, Japan,
is the latest collider experiment in high energy physics. 
SuperKEKB is an accelerator that collides the 7.0 GeV electron beam with the 4.0 GeV positron beam.
The goal of Belle II is to collect 50 \(\text{ab}^{-1}\) of integrated luminosity,
to measure rare decays and CP-violation in heavy quarks and leptons with high precision,
which provides unique probes for new physics beyond the Standard Model.
 The Belle II detector consists of pixel detector (PXD), silicon-strip vertex detector (SVD),
central drift chamber (CDC), time of propagation detector (TOP), aerogel ring-imaging cherenkov detector (ARICH),
electromagnetic calorimeter (ECL) and K-long and muon detector (KLM).

 Our data acquisition (DAQ) system consists of various parts.
Detector front-end electronics (FEE) boards receive triggers and clock signal
from the unified trigger timing distribution (TTD) system with frontend timing switch  (FTSW) modules.
 Then digitized signal from the FEE boards, except for PXD, are transferred to
a unified readout boards called 
common pipelined platform for electronics readout (COPPER) via a 
high speed optical link (Belle2link)\cite{b2}.
The readout PC gathers data from the COPPER boards via Ethernet
and send it to high level trigger (HLT) input server 
where the event building is performed.

Belle II has started operation with full subdetectors in March 2019.
 One of the essential task to ensure stable data-taking at Belle II is to collect and archive various status of Belle II subdetectors.
In this presentation, we present the archiver management system of Belle II, 
including the experimental physics and industrial control system (EPICS) archiver appliance, 
how we secure operation, the type of stored contents in the archiver,
the process of monitoring the incoming stream of information, 
and the use of the archived information.

\section{Hardwares and Backend Software}
    \subsection{Network configuration}
\begin{figure}[htbp]
        \centerline{
            \includegraphics[width=\linewidth]{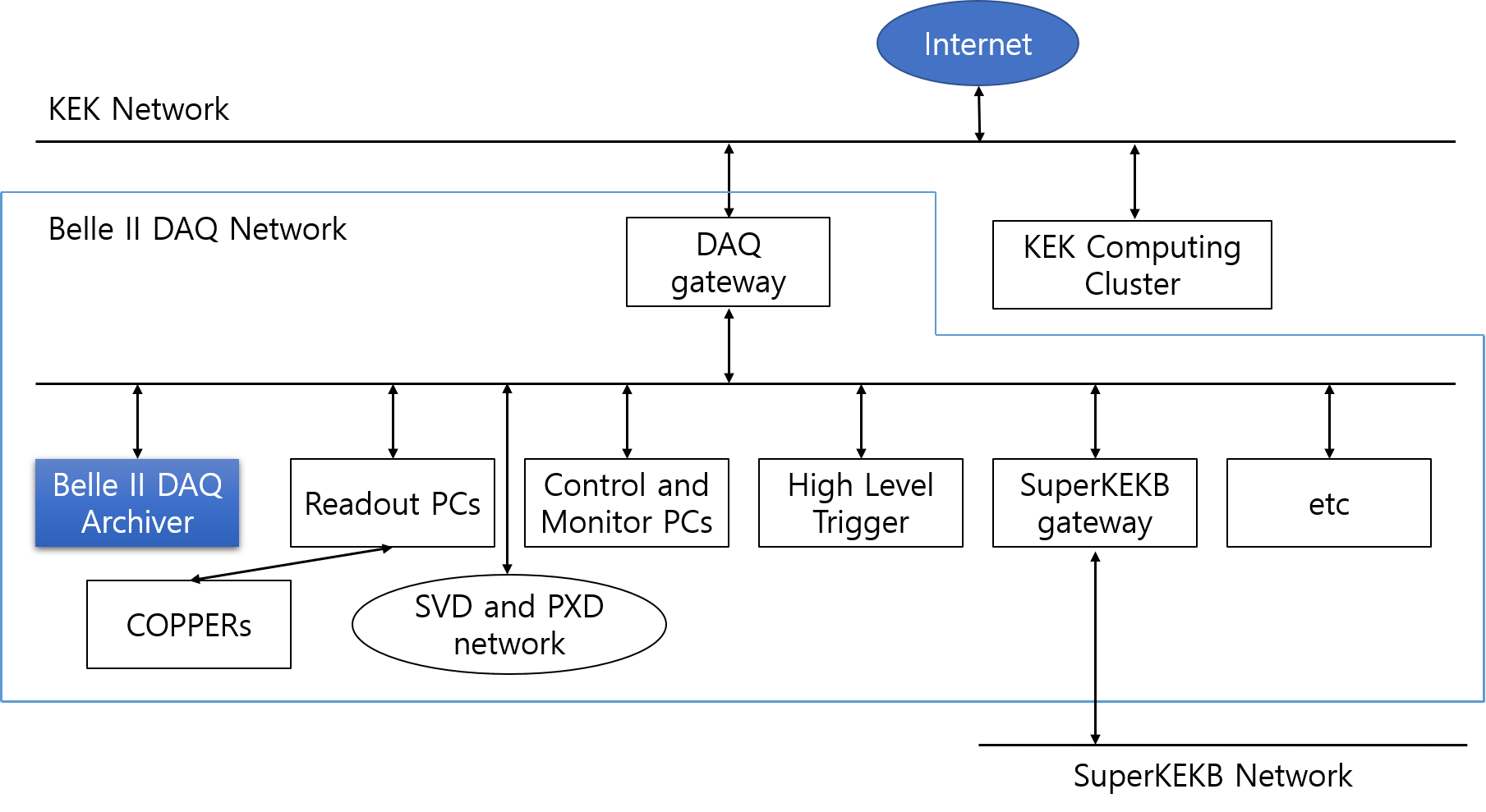}
        }
        \caption{Belle II DAQ network}
        \label{fig_nsm_epics_net}
    \end{figure}

    For each subdetector system, we have dedicated readout system 
    which resides in the Electronic hut (E-hut), located next to the detector.
    Each subdetector has up to ten readout PCs, 
    Each readout PC collects data from COPPERs 
    through dedicated network connections, and also controls COPPERs.  
    Readout PCs also collect status of COPPERs and FEEs
    whose information is collected by COPPERs through Belle2link.
    Belle II DAQ network is accessible 
    only inside the Belle II experimental hall.
    Belle II DAQ network consists of two parallel networks. 
    One is dedicated for slow control system 
    and the other is various purposes including system services,
    interactive operations and file transfer.
    Our EPICS and NSM networks reside inside Belle II DAQ networks.
    KEK computing cluster (KEKCC) can be accessed 
    using a dedicated tunnel from DAQ networks as shown in Fig. \ref{fig_nsm_epics_net}. 

    \subsection{Backend Software}
        To distribute our monitoring information, 
        we use network shared memory 2 (NSM2)\cite{b4} variables and EPICS process variables (PV).
        Because DAQ slow control system transfer its information using NSM2, 
        and monitoring system uses CS-studio which is compatible with EPICS. 
        Network shared memory was developed for Belle Experiment, 
        and upgraded to NSM2 for Belle II experiments.
        NSM2 Variables can be transferred using dedicated NSM2 networks
        inside our DAQ networks,
        but our archiver uses EPICS archiver appliance. 
        EPICS archiver appliance can not receive NSM2 variables.
        So we created nsm2cad which converts nsm variables to EPICS PVs.
        These EPICS PVs can be transferred using EPICS networks inside our DAQ networks.

    \subsection{Archiver}
        Our archiver is based on EPICS archiver appliance.
        EPICS archiver appliance is an implementation of an archiver for
        EPICS control systems that aims to archive millions of PVs.
        Our archiver stores environmental and safety variables, 
        and subdetector status.
        We use MySQL that makes database engine store computed variables 
        and updated regularly in table.  
        We prepared two identical archiver as shown below. 
        \begin{itemize}
            \item CPU : Intel(R) Xeon(R) CPU E3-1230 v6 @ 3.50GHz
            \item RAM : Kingston 2400MHz 16GB x2
            \item STS : 250 GB SSD
            \item MTS : 10 TB HDD
            \item LTS : 10 TB HDD
        \end{itemize}
        The main archiver that stores real-time data,
        and the other archiver for backup purpose.
        PXD and SVD have there own dedicated archiver,
        for their own complex data acquisition systems.

\section{Archiver Services}
    \subsection{Archived information}
        In the archiver system, we store a large number of variables. For example, the temperature of various subdetector components,
        status of water leak sensors, high voltage power supply status, data acquisition status,
        and luminosity information of colliding beams. 
        These variables are collected form subsystems such as the subdetector readout systems, 
        central logging system, data acquisition system 
        and the gateway to the SuperKEKB EPICS network.
        Currently, approximately 26k PVs are archived as shown in Table. \ref{tab_npv}.
        Based on our stress test, 
        The system can stably archived ~80k PVs 
        and can archive 100K PVs at maximum. 
        \begin{table}
            \caption{Number of archived PVs}
            \centerline{
                
                \begin{tabular}{c|c|c}
                        & Number of PVs &  \\                   
                    \hline
                    PXD & 20 & have own archiver\\
                    SVD & 354 & have own archiver\\
                    CDC & 465 & \\
                    TOP & 3065& \\
                    ARICH & 15948& \\
                    ECL & 244 & \\
                    KLM & 1023& \\
                    Trigger & 72 & \\
                    ECL Trigger & 851 & \\
                    DAQ & 210 & \\
                    SuperKEKB & 795 & \\
                    etc & 9 & \\
                \end{tabular}
            }
            \label{tab_npv}
        \end{table}
        
\begin{figure}[htbp]
            \centerline{
                \includegraphics[width=\linewidth]{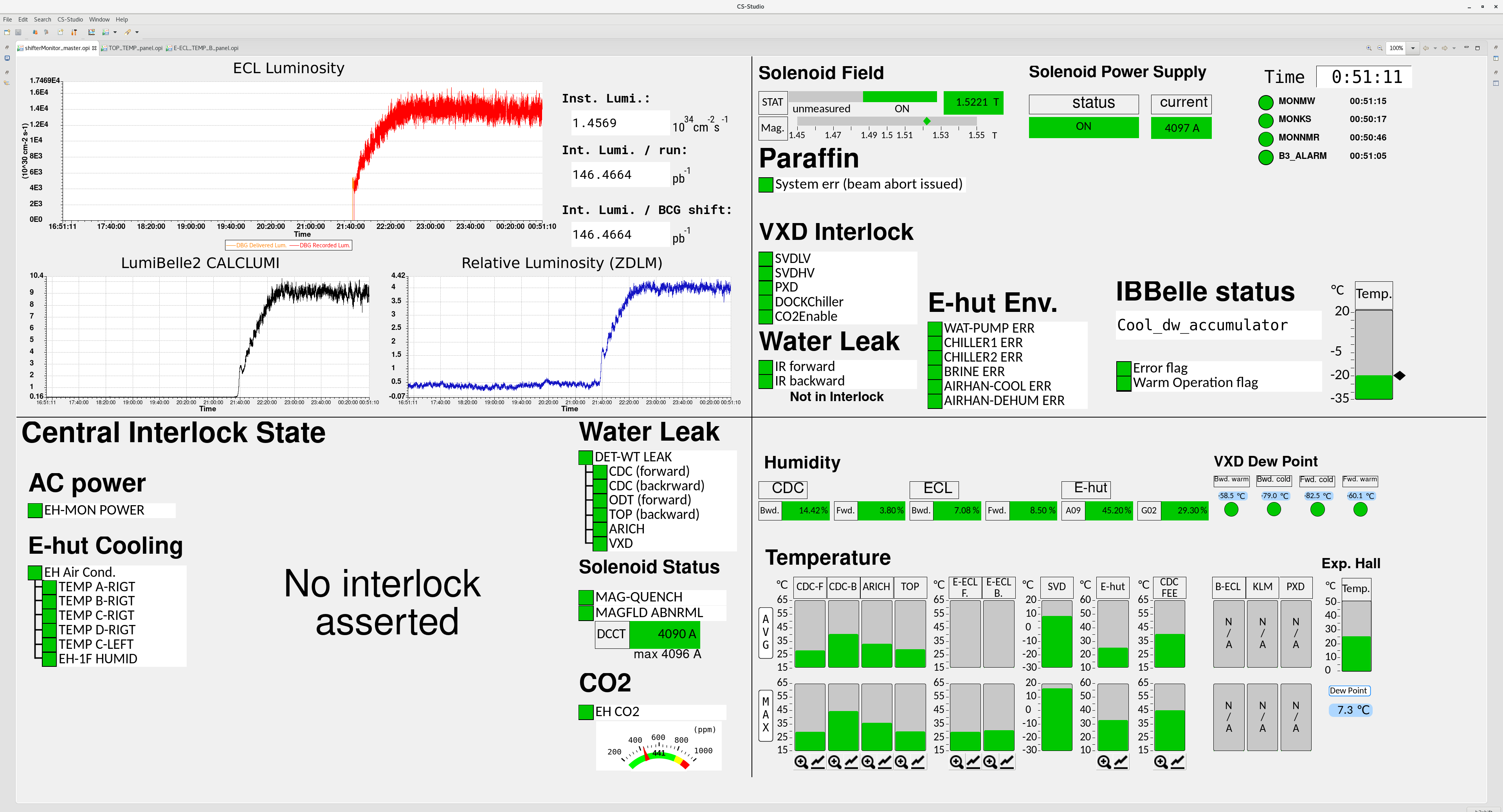}
            }
            \caption{Safety monitoring system using EPICS PVs}
            \label{fig_safetymon}
        \end{figure}

        In our archiver, we store various variables.
        Most of the subdetector groups store variables for safe detector operation such as 
        subdetector temperature water leak status, and humidities.
        We utilize EPICS PV to monitor in Fig. \ref{fig_safetymon}. 
        SVD group stores hardware occupancy. VXD stores radiation.
        CDC group stores high voltage status, FEE status, and gas status.
        TOP group also stores high voltage status, gas status,
        PMT hit rate and FEE status. 
        ARICH group stores low voltage, high voltage status, FEE status,
        cooling water temperature, frontend boards seu status.
        ECL group stores luminosity information and low-level details.
        KLM group stores high voltage status, value of the Data Concentrator (DC) status registers
        and resistive plate chamber (RPC) gas supply status.
        Trigger(TRG) group stores each step of trigger status.
        ECLTRG group stores luminosity information, hitrate, pedestal status,
        noise, temperature, trigger bits, and level-1 trigger information.
        DAQ group stores HV master status, Run control status, HLT event input rate,
        TTD status, event builder (EB) status.
        From superKEKB side, we store MDI flags, information about beams and bunches,
        injection information, Tune parameters, collimator parameters, pressure,
        loss monitor and impact point profile. 
        We use short-term storage (STS), mid-term storage (MTS), 
        and long-term storage (LTS) to store data. 
        STS uses SSD for storage. MTS and LTS use HDD.
        The data are stored in the PB (protocol buffers) format. 
        We make use of the archiver to display information 
        through the CS-studio interface and web browsers 
        for online monitoring purpose.

    \subsection{Archiver Migration and Backup}
\begin{figure}[htbp]
            \centerline{
                \includegraphics[width=0.8\linewidth]{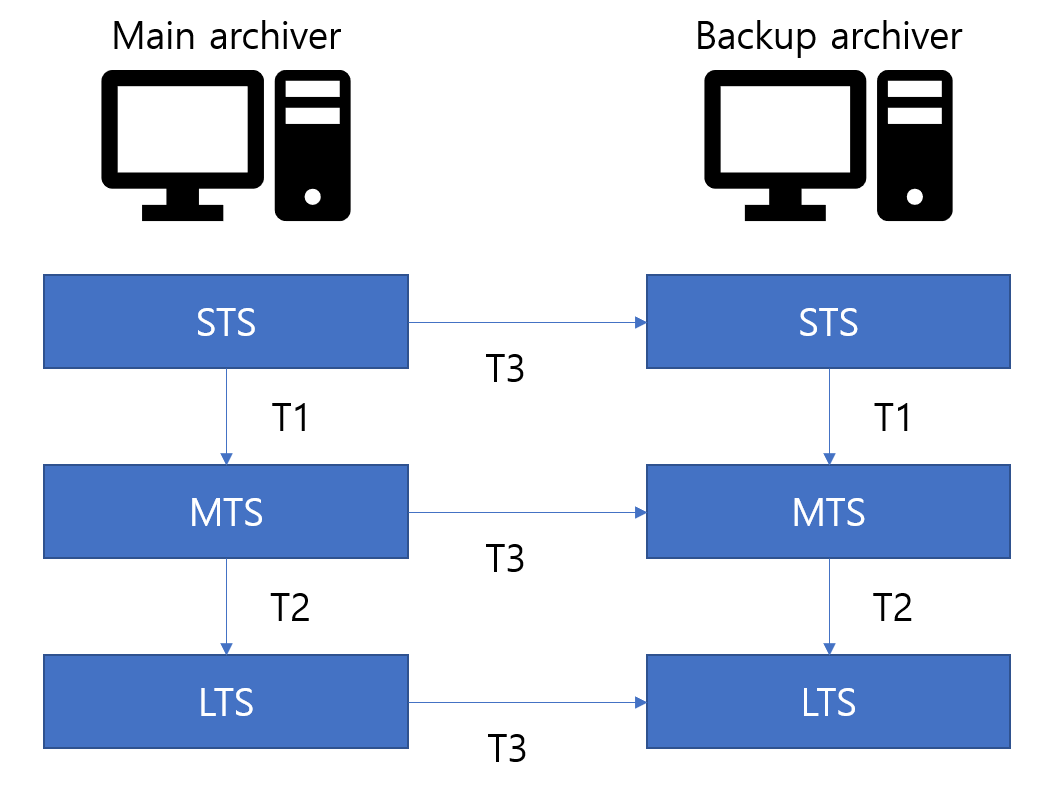}
            }
            \caption{Data transfer and synchronization inside archiver system}
            \label{arch_and_backup}
        \end{figure}

        In our archiver storage, we have STS, MTS, LTS.
        The data stored in STS are merged to MTS in 3 days (T1).
        MTS data are transferred to LTS in a week (T2).
        For backup purpose, we have an identical archiver prepared.
        We synchronize main archiver and backup archiver every 20 minutes(T3).
        In case of emergency, backup archiver performs same operations
        as main archiver as shown Fig. \ref{arch_and_backup}.

    \subsection{Offline use of archived data}
\begin{figure}[htbp]
            \centerline{
                \includegraphics[width=0.8\linewidth]{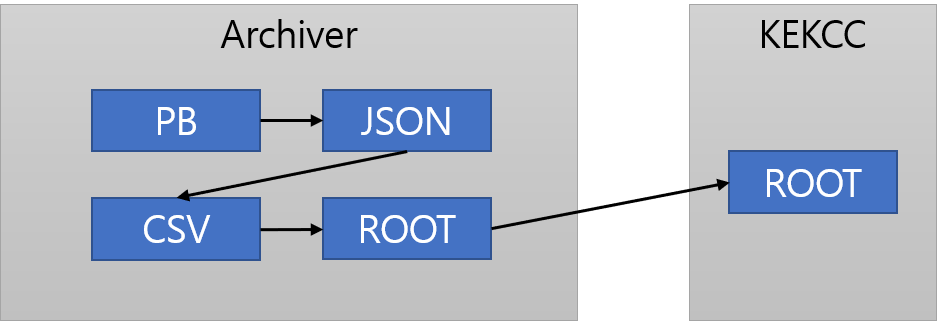}
            }
            \caption{Daily PB to ROOT conversion procedure}
            \label{pbtoroot}
        \end{figure}

        EPICS archiver appliance provides retrieval webapp 
        and supports offline download. 
        Through retrieval webapp, we can easily check status of the PVs.
        Offline download is one of the easiest way to access the archiver data to use in a batch process.
        However, the function only supports archiver specific format called PB,
        which is not analysis friendly.
        Moreover, these data can only be accessible in Belle II DAQ network.
        We developed a chain of conversion programs to convert the PB data 
        into the ROOT\cite{b3} format by combining existing tools.
        As archiver data are available only inside the closed data-acquisition network 
        which is not accessible from offline computers, 
        we make a daily conversion and file transfer 
        to the offline computing system at the KEK (KEKCC).
        First we convert PB to JSON (JavaScript Object Notation) format using function of EPICS archiver appliance, 
        and we convert JSON to CSV using a java program.
        Then we convert CSV file to ROOT format which is the data format adopted 
        by the Belle II analysis software framework as shown in Fig. \ref{pbtoroot}.
        We transfer these converted ROOT files to KEKCC.
        We convert our monthly PB file in LTS and transfer it to KEKCC.

\section{Managing archived variables list}
    As our detector have many variables, we need to check the status of the variables,
    such as connectivity, consistency and list of variables for each subdetectors.
    To check connectivity we made short shell and python scripts 
    which makes a list of disconnected variables. 
    Every morning, archiver experts receive a list of disconnected PVs 
    in an e-mail from the archiver.
    Archiver experts then notify the subdetector experts about the disconnected PVs.
    We also have PV list management. As a part of the collaborative service of Belle II,
    we use a wiki system of Atlassian Confluence. 
    We provide wiki page of a list of archived PVs.
    A confluence wiki page is usually designed for human readers, 
    but we keep a certain format to be able to parse using python script.
    The script compares list of archived PV in archiver and the list on the confluence page.
    Based on the output of this script, archiver expert notifies subdetector experts.
    This procedure can eventually be automated.

\section{Monitoring Archiver}
    During the operation, we use online chat system called Rocket.Chat to communicate 
    among practically everybody who participate in the detector operation, 
    including the local and remote shifter, subdetector and DAQ experts, 
    and Belle II commissioning groups.
    We use Rocket.Chat for fast notification and e-mail 
    for slow notification.
    \subsection{Archiver status}
        To monitor archiving status, we made a dedicated PV. 
        This PV is updated every ten seconds, 
        so we can monitor file saving status by checking stored file information.  
        Our crontab job checks every minute for STS, MTS, LTS. 
        Upon failure, we will get a e-mail and Rocket.Chat notification.
        For STS, if the storage status does not update for more than ten minutes, 
        we will get notified first, and get additional notification every 2 hours.
        For MTS, if the storage status does not update for an hour, we receive notification, 
        and if the failure sustains we receive further notification every 2 hours.
        For LTS, if the storage status does not update for a day, 
        we will receive first notification.
        If the failure lasts, we get further notification every 4 hours. 
    \subsection{Retrieval status}
\begin{figure}[htbp]
            \centerline{
                \includegraphics[width=\linewidth]{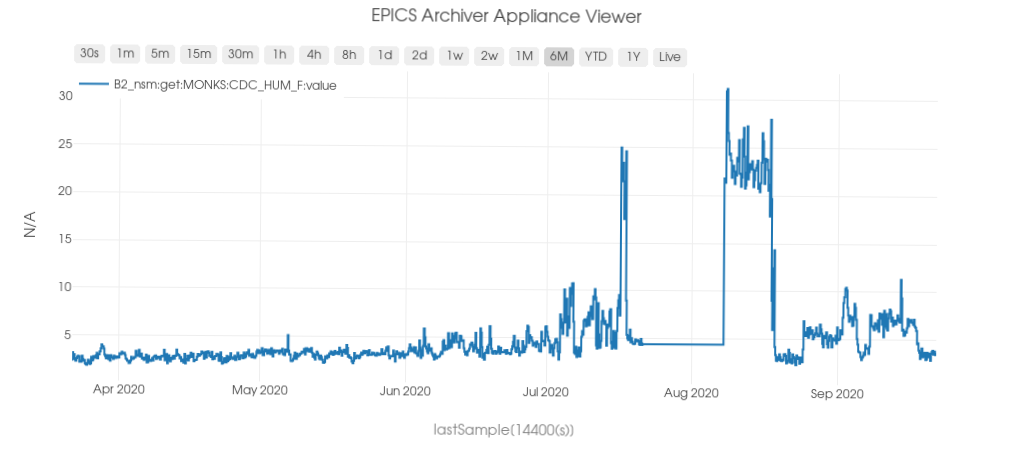}
            }
            \caption{Archived CDC humidity variable distributed using retrieval webapp}
            \label{fig_retrieval_ex}
        \end{figure}

        Our retrieval webapp is one of the parts of archiver data distribution system.
        We found in some cases, when a flood of data requests exceeds the capability of the archiver, 
        the retrieval webapp, in Fig. \ref{fig_retrieval_ex}, stops responding.
        In order to detect such situation, we check every ten minutes 
        whether the retrieval webapp is working or not
        by requesting data for past one hour from another PC.
        If retrieval webapp does not respond, 
        our alarm system sends an alarm to Rocket.Chat every ten minutes. 

    \subsection{Fault tolerance}
        In case of a hardware failure of the archiver, 
        the archiver can be switched to the backup system simply 
        by manually switching the IP address.  
        In this case, we might lose up to 20 minutes plus 
        the time for the switching operation of archived data, 
        since the backup system is synchronized every 20 minutes.
        The data files copied to KEKCC can also be served as a backup.

\section{Summary}
     To secure stable data taking of archiver system for Belle II detector operation, 
    we monitor archived variable status, manage archived variables list, 
    and monitor the archiver itself.
    With these efforts we can ensure that, 
    and we would not lose more than 20 minutes of data even in the worst situation.
    We are operating our archiver from April, 2019 successfully.

\appendices


\end{document}